\documentclass[a4paper,twocolumn]{article}

\usepackage{graphicx}
\usepackage{hyperref}

\begin{document}
\title{Is the P300 Speller Independent?}
\author{Stefan Frenzel and Elke Neubert\\~\\Institute of Mathematics and Informatics, University of Greifswald\\ Greifswald, Germany\\ \small{ \texttt{frenzel@physik.uni-greifswald.de}}}
\date{June 15, 2010}    
\twocolumn[
\begin{@twocolumnfalse}
\maketitle
\begin{abstract}The P300 speller is being considered as an independent brain--computer interface. That means it measures the user's intent, and does not require the user to move any muscles. In particular it should not require eye fixation of the desired character. However, it has been shown that posterior electrodes provide significant discriminative information, which is likely related to visual processing. These findings imply the need for studies controlling the effect of eye movements. In experiments with a $3\times3$ character matrix, attention and eye fixation was directed to different characters. In the event--related potentials, a P300 occurred for the attended character, and N200 was seen for the trials showing the focussed character. It occurred at posterior sites, reaching its peak at 200ms after stimulus onset. The results suggest that gaze direction plays an important role in P300 speller paradigm. By controlling gaze direction it is possible to separate voluntary and involuntary EEG responses to the highlighting of characters. \\
\end{abstract}
\end{@twocolumnfalse}
]
\section*{Introduction}Brain--computer interfaces (BCIs) aim to provide new channels for man--machine communcation and control using brain signals. Numerous studies in the past showed that electroencephalography (EEG) is suitable for this purpose \cite{Vidal1977,Farwell1988,Wolpaw1991,Sutter1992,Pfurtscheller1993,Birbaumer1999}. EEG--based BCI systems are non--invasive and comparatively easy to set up. They are especially suited for real--time applications due to their high temporal resolution. The P300 speller is one type of BCI first introduced by Donchin and Farwell \cite{Farwell1988}. Let us review the basic idea behind it. The subject sits in front of a screen presenting a matrix of characters. Rows and columns of characters are highlighted in random order, while the subject focuses his attention to one desired (target) character. This is usually done by performing a mental count of the number of times the target character is highlighted. Whenever the target character is highlighted a P300 is elicited. P300 (P300b) is a component of the event--related potential (ERP) typically appearing as a positivity at electrodes covering the parietal lobe 300ms to 400ms after stimulus onset \cite{Sutton1965,PolichKok1995,PatelAzzam2005}. It is an endogenous (voluntary) component,  mainly reflecting mechanisms of attention allocation and immediate memory \cite{PolichKok1995, PatelAzzam2005}. This allows for inferring the target character from the EEG data by various statistical methods \cite{Krusienski2006,Bandt2009}. The P300 speller needs little training and has been proven suitable for most persons \cite{Guger2009}. It has been sucessfully implemented as a support for persons with physical disabilities, e.g. locked--in amyotrophic lateral sclerosis (ALS) patients \cite{Vaughan2006}. 

According to \cite{Wolpaw2002}, a BCI is called independent if it does not make use of the brain's normal output pathways of peripheral nerves and muscles. It measures the users intention without requiring any movements, e.g. eye movements. P300 reflects cognitive processes related to the perception of the target character. Since the target character can be attended covertly without fixating upon it \cite{Posner1980} the P300 speller is in general considered as an independent BCI. On the other hand, it turned out that electrodes located over posterior lobe provide discriminative information, which is not due to P300 \cite{BlankertzCurio2003,Kaper2004}. Whereas P300 is usually measured at electrode Pz \cite{Sutton1965,Farwell1988}, adding posterior electrodes turned out to improve the classification performance of the P300 speller \cite{Kaper2004,Krusienski2008,Hoffmann2008}.
Waveform analysis shows a negative deflection preceding P300, approximately 200ms after stimulus onset \cite{Krusienski2008,Hoffmann2008,Hong2009}, which in the sequel is referred to as N200. This component is well known to be elicited in visual oddball experiments. Several sub--components were found which reflect both voluntary and involuntary processing of visual stimuli \cite{Makeig1999, PazoAlvarez2003, PatelAzzam2005}. Its role within the speller paradigm is not clear yet. It has been hypothesized that N200 is related to visual processing of the fixated character \cite{BlankertzCurio2003,Kaper2004,Krusienski2008}. Transient visual evoked potentials (VEPs) due to highlighting of characters may contain components with peak latencies up to approximately 200ms \cite{Regan1972}. If indeed N200 is due to involuntary visual processing, we expect it to be strongly correlated with the fixated character. On the other hand, covert attention has been shown to modulate ERP components at posterior sites as early as 160ms \cite{Nobre2000}. 
If N200 mainly reflects voluntary processing, it should be correlated with the attended character. In this case the standard P300 speller can be regarded as independent. However, it seems that P300 speller studies do not control the effect of eye movements. Therefore we performed exeriments where the attended character differs from the fixated one. Subjects had to fixate one character while attending another one covertly.

Donchin and Farwell originally introduced the P300 speller using a $6\times6$ matrix of characters, with rows and columns flashing successively at high presentation rate \cite{Farwell1988}. For the purpose of the study there are few reasons to alter the standard settings. First of all, covert attention to single characters may be difficult when using a large character matrix. 
We therefore used a $3\times3$ matrix, allowing for larger character size. Since P300 amplitude varies indirectly with the a priori probability of the target stimulus \cite{Duncan-JohnsonDonchin1977}, single characters instead of rows and columns are highlighted. Characters were dark grey on light grey background and were set to black during the highlighting. It has been reported that this setup may be more convenient than the original white--on--black setting \cite{Salvaris2009}. Another problem comes with usage of small interstimulus intervals in the standard P300 speller. It was shown to introduce overlap and refractory effects of the ERPs \cite{MartensEtAl2009}. We avoid such problems by using longer interstimulus intervals than the standard one.
 
\section*{Methods}
\subsection*{Participants}
Ten healthy volunteers (2 male, 8 female) with normal vision participated in the present study. All subjects except one had no previous experience with BCI--related experiments. Subjects gave written informed consent after the experiment was explained to them. 
\begin{figure}
\centering
\includegraphics[width=2.9in]{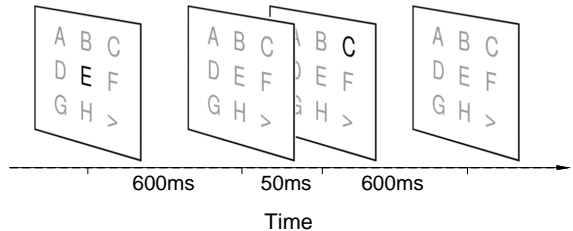}
\caption{{Schematic representation of the stimulus paradigm.} Single characters were highlighted for 600ms followed by a randomised break lasting up to 50ms.}
\label{fig:stimulus_schema}
\end{figure}
\subsection*{Experimental Setup}
The subjects sat in front of a 19 inch screen in a distance of about 50cm. They were instructed to relax and avoid blinking during the recordings as far as possible. Each single--trial lasted 600ms and one single character was highlighted during this time. The sequence of characters was randomised, with each character being highlighted the same number of times, and a total number of trials lying between 450 and 477. Between two consecutive trials there was a randomised break lasting up to 50ms, where no character was highlighted (see Figure \ref{fig:stimulus_schema}). Hence the interstimulus interval varied between 600ms and 650ms. The experiment consisted of two parts, with a short break in between. \\ 
\textbf{(I) Combined attention and fixation.} In the first part of the experiment subjects had to fixate the target character ``E'' in the center of the screen. They were instructed to mentally count the number of times it is highlighted. Afterwards they were asked for the count and given feedback on the correctness.\\
 \textbf{(II) Separated attention and fixation.} In the second part subjects still had to fixate the character ``E'', but count the new target character ``B'' instead. As in the first session they were instructed to perform a mental count of the number of times the target character is highlighted. They were asked for the count and given feedback on the correctness. The total number of trials was different from the one in the first session.\\ 
Afterwards subjects used the P300 speller in online mode using the data from the first experiment as training for linear discriminant analysis \cite{Bandt2009}. This part will not be discussed here.
\begin{figure*}[t!!]
\begin{center}
\includegraphics[width=6.1in]{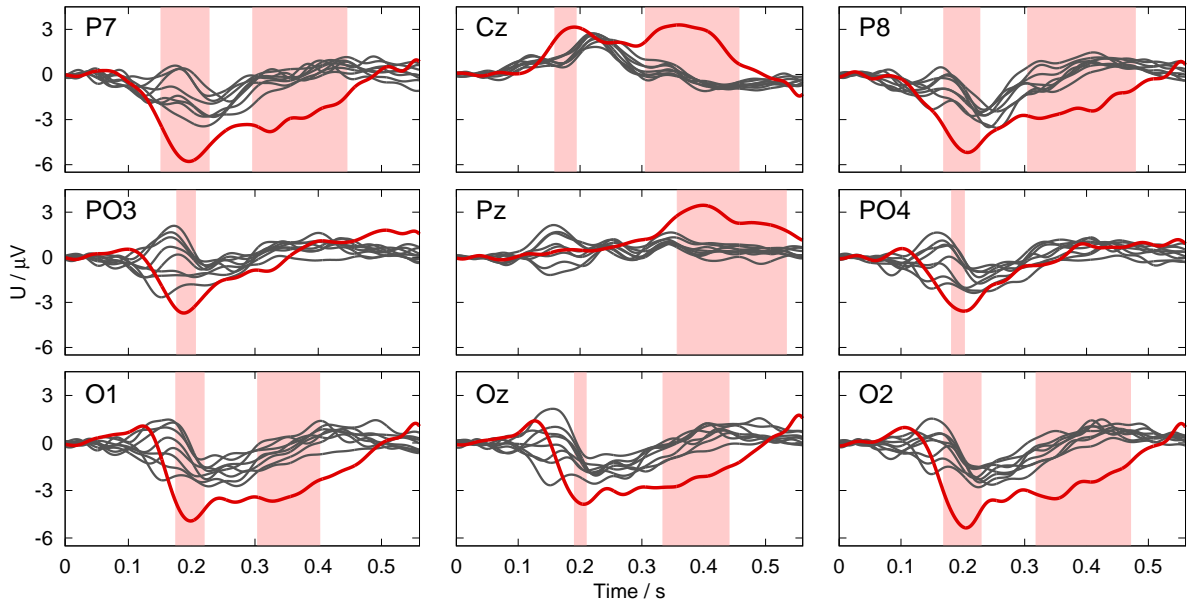}
\end{center}
\caption{{ERP waveforms for combined attention and fixation.} Grand average ERPs for attended/fixated character (red), and the eight remaining ones (grey). Time points where t--tests revealed significant differences between red and each grey curve were shaded red ($\alpha=0.1\%$).}
\label{fig:grandaverage_part1}
\end{figure*} 
\subsection*{Matrix style}
We used a $3\times3$ matrix of characters. To access all characters of the alphabet in the online session after the experiments, we used additional symbols for switching the character set located at the lower right and left edge (see Figure \ref{fig:stimulus_schema}).
Characters were dark grey (rgb 128, 128, 128) on light grey background (rgb 200, 208, 212) and were set to black during the highlighting.
\subsection*{Data collection and preprocessing}
The data were recorded using a Biosemi ActiveTwo system with 32 electrodes placed at positions of the modified 10--20 system and sampling rate of 2048Hz. They were referenced by subtracting the mean over all electrodes for each time point. To normalize data in time we subtracted the mean of the first 50ms from each single trial. We applied a 5th order Chebyshev type II bandpass filter with edge frequencies 0.1Hz and 48Hz, and stopband ripple 50dB down from passband. To avoid phase shifts the filter was applied both forward and backward in time.
Trials containing blinks were rejected after visual inspection of channel Fp1. The relative number of trials rejected was smaller than $12\%$ for all subjects, resulting in a net number of 399 to 473 trials for single subjects. 
\section*{Results}
\subsection*{Combined attention and fixation}
\begin{figure}
\begin{center}
\includegraphics[width=2.8in]{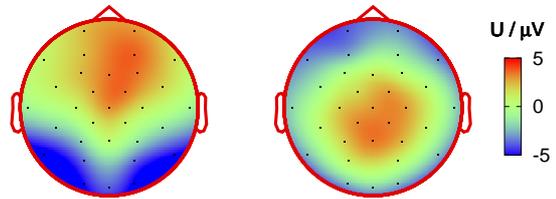}
\end{center}
\caption{{ERP topographies of N200 and P300 for combined attention and fixation.} The grand average ERP to the target character contained two major components, one posterior negativity at 200ms (left), and a late positivity at parietal sites, reaching a peak at 400ms (right).}
\label{fig:topopgraphic_maps_EE}
\end{figure}

In the first part of the experiment subjects had to fixate and count the target character ``E'' while neglecting all others. Figure \ref{fig:grandaverage_part1} shows the grand average ERPs for all trials presenting the target character (red curve), and trials presenting the eight non--target ones (grey curves). To examine significant differences among them, we performed t--tests for equality of target and each non--target ERP. Time points where all eight null hypotheses were rejected at $0.1\%$ level were shaded red. 
The ERP of the target character contained a P300 component, reaching its peak 400ms after stimlus onset with an amplitude of approximately $4\mu\rm{V}$ at electrode Pz. P300 waveform had a rather broad shape, starting from 300ms and lasting until 600ms. Figure \ref{fig:topopgraphic_maps_EE} shows the topographic map of the ERP for the target character at 400ms. It shows the typical spatial distribution with a positivity at parieto--central electrodes and declining to the periphery \cite{Sutton1965,Farwell1988}.  
\begin{figure*}[t!!]
\begin{center}
\includegraphics[width=6.1in]{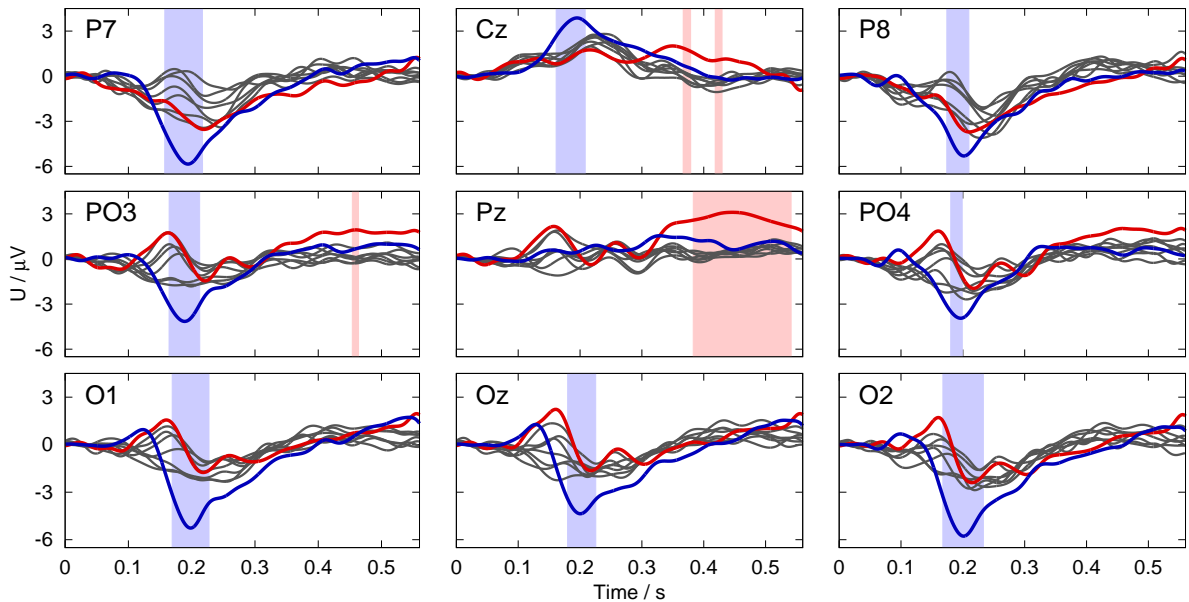}
\end{center}
\caption{{ERP waveforms for separated attention and fixation.} Grand average ERPs for attended (red), fixated (blue), and the seven remaining characters (grey). Time points where t--tests revealed significant differences between red/blue curve and each remaining one were shaded red/blue ($\alpha=0.1\%$). }
\label{fig:grandaverage_part2}
\end{figure*}
Besides this, the data also contained discriminantive information at posterior electrodes before 300ms (see Figure \ref{fig:grandaverage_part1}). In agreement to the findings in \cite{Krusienski2008,Hoffmann2008,Hong2009} there was a N200 appearing as a negative deflection at posterior sites at both hemispheres symmetrically. N200 reached a first peak 200ms after stimulus onset with an amplitude of approximately $-6  \mu\rm{V}$ at P7. The corresponding topographic map is shown in Figure \ref{fig:topopgraphic_maps_EE}. The most complete picture of N200 within the P300 speller has been given in \cite{Hong2009} during a comparison with a new type of speller paradigm.  Although using a $6\times6$ character matrix, they find similar topography and latency.

Since the fixated character coincided with the attended one, we can not understand the nature of N200 from the first part of the experiment. It could be both related to involuntary processing due to eye fixation and to voluntary processing because subjects were attending the target character while performing the mental count. We therefore separated both in the second experiment.
\subsection*{Separated attention and fixation}
\begin{figure}
\begin{center}
\includegraphics[width=2.8in]{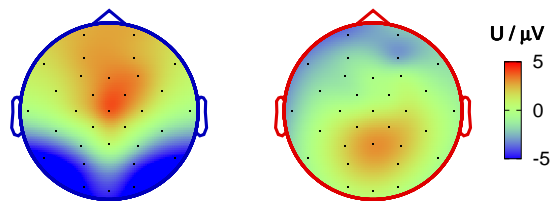}
\end{center}
\caption{{ERP topographies of N200 and P300 for separated attention and fixation.} As in the first part of the experiment, the grand average ERP to the attended target character contained a late positivity at parietal sites, reaching a peak at 450ms (right). In contrast, N200 occurred for the fixated character around 200ms after stimulus onset (left).}
\label{fig:topopgraphic_maps_EB}
\end{figure}

In the second part of the experiment subjects still had to fixate the character ``E'', but now count ``B'' instead. This allowed for separating involuntary and voluntary processing by analyzing the ERPs to the attended target character and the fixated one, respectively. Figure \ref{fig:grandaverage_part2} shows the grand average over trials presenting the fixated character (blue curve), the attended one (red curve), and the remaining seven non--target ones (grey curves). Once again, we performed t--tests for equality of ERPs to fixated and each non--fixated character as well as attended and each non--attended character. Time points where all eight null hypotheses were rejected at the $0.1\%$ level were shaded blue and red respectively. 
It is well known that visuospatial attention can be directed to non--fixated locations in the visual field \cite{Posner1980}. Hence we expected P300 appearing for the attended target character ``B''. Indeed there was a late positivity with an amplitude of approximately $3 \mu\rm{V}$ at electrode Pz, as can be seen from the red curve in Figure \ref{fig:grandaverage_part2}. Figure \ref{fig:topopgraphic_maps_EB} shows the corresponding topographic map at 450ms. So although the target character was not fixated, it elicited a P300 due to covert attention when subjects were performing the mental count. 

If N200 also represents voluntary processing similar to P300, it should  for the attended character and not the fixated one. Our results indicate, that this is not the case. In contrast to the first experiment, N200 occurred for the fixated character and not the attended one (blue curve in Figure \ref{fig:grandaverage_part2}). Note the remarkable similarity with the ERP to the target character in the first experiment within the first 250ms (red curve in Figure \ref{fig:grandaverage_part1}). Indeed, except for electrodes C3 and AF4, t--tests revealed no significant differences between both within the first 280ms ($\alpha=0.1\%$). There was no significant difference before 210ms for all electrodes.
\section*{Discussion}
There have been speculations on the role of gaze direction in P300 speller BCI \cite{BlankertzCurio2003,Kaper2004,Krusienski2008,Hoffmann2008}. However, to our best knowledge, so far no study controls the effect of eye movements in P300 speller paradigm. Hence we performed experiments where the attended character was either identical with the fixated one or where it differed from the last. We investigated the corresponding grand average ERPs and found two major components. As expected there was a P300 occurring for the attended target character \cite{Sutton1965,Farwell1988}. It was elicited also in the second part of the experiment, where the target character was not fixated at the same time. In contrast, N200 occurred for the fixated character and not the attended one in both parts. This strongly suggest that N200 reflects involuntary processing of the fixated character. Moreover the ERP to the fixated character was virtually identical in both parts of the experiments up to 280ms after stimulus onset. These findings have consequences for the construction of independent P300 spellers \cite{Wolpaw2002}. In general, if the user fixates the target character during the training phase, automatic feature extraction methods will not lead to an independent BCI.

By controlling gaze direction it is possible to separate voluntary and involuntary EEG responses to the highlighting of characters. This finding leads to new interesting questions related to the P300 speller. For instance, how does the use of covert attention only affects the performance of the P300 speller? Is it still possible for most persons to use it at reasonable speed? These are questions which should be adressed in further studies.
\section*{Acknowledgments}
The authors would like to thank Christoph Bandt for fruitful discussions.
\bibliography{bci}
\bibliographystyle{unsrt}
\end{document}